\input phyzzx.tex
\tolerance=1000
\voffset=-0.0cm
\hoffset=0.7cm
\sequentialequations
\def\rl{\rightline}

\def\t1{{\tilde 1}}

\REF{\DRT}{M. Dine, L.Randall and S. Thomas, Phys. Rev. Lett. {\bf 75} (1995)
398, hep-ph/9503303; Nucl. Phys. {\bf B458} (1996) 589.}
\REF{\RT}{L. Randall and S. Thomas, Nucl. phys. {\bf B449} (1995) 229,
hep-ph/9407248;
B. de Carlos,J. A. Casas, F. Quevedo and E. Roulet, Phys.Lett. {\bf B318}
(1993) 447, hep-ph/9308325.}
\REF{\LI}{A. Linde, Phys. Rev. {\bf D53} (1996) 4129, hep-ph/9601083.}
\REF{\DIM}{N. Arkani-Hamed, S. Dimopoulos and G. Dvali,  Phys. Lett.
{\bf B429} (1998) 263, hep-ph/9803315;
I.~Antoniadis, N.~Arkani-Hamed, S.~Dimopoulos and G.~Dvali,
Phys. Lett. {\bf B436}, 257 (1998), hep-ph/9804398.}
\REF{\OLD}{I. Antoniadis, Phys. Lett. {\bf B246} (1990) 377; J. Lykken,
 Phys. Rev. {\bf D54} (1996) 3693, hep-th/9603133; I. Antoniadis and K.
Benakli, Phys. Lett. {\bf B326} (1994) 69;
I. Antoniadis, K. Benakli and M. Quiros, Nucl. Phys. {\bf B331} (1994) 313.}
\REF{\LYT}{ D.H. Lyth,  Phys. Lett. {\bf B448} (1999) 191, hep-ph/9810320.}
\REF{\KL}{ N. Kaloper and A. Linde, hep-th/9811141.}
\REF{\NIM}{N. Arkani-Hamed, S. Dimopoulos, N. Kaloper and J. March-Russell,
hep-ph/9903224.}
\REF{\BEN}{K. Benakli, hep-ph/9809582; K. Benakli and S. Davidson,
hep-ph/9810280.}
\REF{\MR}{ M. Maggiore and A. Riotto, hep-th/9811089.}
\REF{\DIE}{K.R. Dienes, E. Dudas, T. Ghergetta, A. Riotto, hep-ph/9809406.}
\REF{\DT}{G. Dvali and S.-H. Tye, hep-ph/9812483.}
\REF{\CGT}{C. Csaki, M. Graesser and J. Terning, hep-ph/9903319.}
\REF{\CLI}{J. Cline, hep-ph/9904495.}
\REF{\RIO}{A. Riotto, hep-ph/9904485.}
\REF{\DVA}{G. Dvali, hep-ph/9905204.}
\REF{\EH}{E. Halyo, hep-ph/9905244.}
\REF{\OTH}{A. Lukas, B. Ovrut, K. Stelle and D. Waldram, hep-th/9803235;
L. Randall and R. Sundrum, hep-ph/9605221; hep-th/9906064; T.
Nihei,hep-ph/9905487; N.Kaloper, hep-th/9905210; P. Binetruy, C. Deffayet and
D. langlois, hep-th/9905012; C. Csaki, M. Graesser, C. Kolda and J. Terning,
hep-ph/9906513.}
\REF{\HBD}{E. Halyo, {\it Phys. Lett.} {\bf B387}  (1996) 43, hep-ph/9606423;
P. Binetruy and G. Dvali, {\it Phys. Lett.} {\bf B388}  (1996) 241,
hep-ph/9606342; E. Halyo, Phys. Lett. {\bf B454} (1999) 223; hep-ph/9901302.}
\REF{\CUL}{S. Cullen and M. Perelstein, hep-ph/9903422.}
\REF{\AFIV}{G. Aldazabal, A. Font, L. Ibanez and G. Violero, hep-th/9804026.}
\REF{\IMR}{L. Ibanez,  C. Munoz and S. Rigolin, hep-ph/9812397.}
\REF{\NIL}{Z. Lalak, S. Lavignac and H. P. Nilles, hep-th/9903160; I.
Buchbinder, M.Cvetic,
A. Yu. Petrov, hep-th/9906141.}
\REF{\LIN}{A. Linde, Phys. Lett. {\bf B259} (1991)38; Phys. Rev. {\bf D49}
(1994) 748.}
\REF{\EXP}{N. Arkani-Hamed, S. Dimopoulos and G. Dvali,  hep-ph/9807344.}
\REF{\HAL}{G. Shiu, R. Schrock and S. H. Tye, hep-ph/9904262; E. Halyo,
hep-ph/9904432.}

\singlespace
\rl{SU-ITP-99-33}
\rl{hep-ph/9907223}
\rl{\today}
%\rl{T}
\pagenumber=0
\normalspace
\medskip
\bigskip
\titlestyle{\bf{ Late D--term Inflation and the Cosmological Moduli Problem in
TeV Scale Strings}}
\smallskip
\author{ Edi Halyo{\footnote*{e--mail address: halyo@dormouse.stanford.edu}}}
\smallskip
\centerline {Department of Physics}
\centerline{Stanford University}
\centerline {Stanford, CA 94305}
\smallskip
\vskip 2 cm
\titlestyle{\bf ABSTRACT}

We show that a short period of late D--term inflation can solve the
cosmological moduli (radion) problem of  (asymmetric) inflation at the TeV
scale. Late inflation happens after the large compact dimensions are stabilized
which is crucial for obtaining the extremely small Hubble constant and inflaton
mass required.

\singlespace
\vskip 0.5cm
\endpage
\normalspace

\centerline{\bf 1. Introduction}
\medskip

A generic problem in late cosmology of supersymmetric theories with exact flat
directions (or moduli) is the cosmological moduli problem[\DRT]. An important
example is superstring theory
in which there are many moduli such as the dilaton, the untwisted and twisted
moduli which parametrize different string vacua. Another example is radii of
compact dimensions in gravitational theories with more than four dimensions.
(Actualy these are related to the dilaton and untwisted moduli of string
theory.)
These moduli are massless to all orders in perturbation theory and obtain
masses only by nonperturbative effects which break supersymmetry. These effects
lift the flat directions and the moduli get a potential with a minimum.
As a result, the moduli start to roll down from their initial value and begin
to oscillate about their minimum. Since the initial value of the moduli fields
are arbitrary, there may be a large energy density stored in the moduli fields.
The energy density stored in radiation decreases as $a(t)^{-4}$
compared to $a(t)^{-3}$ for moduli ($a(t)$ is the scale factor of the
universe), after some time the moduli energy density dominates the universe.
This is unacceptable since it will modify the successful predictions of
nucleosynthesis and later stages of Big Bang cosmology.

The cosmological moduli problem persists in models of inflationary cosmology.
The only difference is that after inflation and reheating of the universe, the
moduli start rolling down
to their true minimum from their minimum during inflation instead of an
arbitrary initial value. Thus, even though immediately after inflation the
universe is radiation dominated after some time there is a danger that the
moduli energy density will take over.
An elegant solution to the cosmological moduli problem is late inflation, i.e.
a second stage of inflation that occurs after
the moduli start to oscillate about their minimum[\RT]. (For another solution
to the moduli problem see ref. [\LI].)
 If late inflation results in $N$ e--foldings then the moduli energy density is
diluted by a factor of $e^{-3N}$ so that for large enough $N$ the problem is
solved. In supersymmetric theories the moduli masses are of $O(TeV)$ and
therefore one needs inflation around or below this scale.

Recently, a new scenario with (two or more) large internal dimensions and TeV
scale strings has been proposed[\DIM,\OLD].
In this scenario
our world lives on a brane on which there are matter and gauge interactions
whereas gravity lives only in the bulk. The weakness of gravity with respect to
the gauge interactions is a result of the large internal dimensions. It is
clear that this scenario requires a revision of cosmology above the
nucleosynthesis scale, $\sim MeV$.
In particular, inflation has to be realized in a new way.
There have been a number of works related to cosmology and inflation in the
presence of large
compact dimensions[\LYT-\EH] and of branes embedded in higher dimensional
spaces[\OTH]. It is easy to see that the cosmological moduli problem
is even more severe in these new inflationary models due to the very low
energies involved. In this letter, we consider the case of two large dimensions
with a string scale (or six dimensional Planck scale) of $M_s \sim
50~TeV$[\CUL]. However, all our results can be easily generalized to the cases
with more than two large dimensions.

In [\NIM], it was shown that in asymmetric inflation there is a radion problem.
The radii of the compact dimensions have very small masses of $O(10^{-9}~GeV)$
and their energy density can dominate the universe at late times. Similar
problems arise in other inflationary models at the TeV scale (such as D--term
inflation[\EH] or brane inflation[\DT])
which generically have moduli with similar masses.
The moduli problem can be solved by late inflation as mentioned above but now
one needs inflation with a
Hubble constant of at least $H< 10^{-9}~GeV$. (In fact, we will see that there
is a much stronger bound on $H$.)
Moreover, the inflaton mass must be smaller than $H$ for inflation to occur and
this is difficult to achieve. With such a light inflaton
it is very hard to reheat the universe to above the nucleosynthesis scale.

In this letter, we show that late D--term inflation[\HBD] can provide a
solution to the cosmological moduli (or radion) problem. Since the Hubble
constant is very small late D--term inflation occurs after
the large compact dimensions are stabilized. Then the large dimensions give
rise to the extremely small gauge coupling (for the anomalous $U(1)$) which is
crucial for obtaining the extremely small Hubble constant and inflaton mass
required. In the next section we show how late D--term inflation solves the
radion problem of early asymmetric inflation. Section 3 contains our
conclusions and a discussion of our results.

\bigskip
\centerline{\bf 2. The Radion Problem in Early Asymmetric Inflation}
\medskip

In this section, we assume that the horizon, monopole and flatness problems are
solved by an early phase of asymmetric inflation
at a Hubble constant $H \sim M_s \sim 5 \times 10^{4} ~GeV$ in a theory with
TeV scale gravity[\NIM]. Asymmetric inflation occurs when the size of all the
compact dimensions are of $O(M_s^{-1})$. During asymmetric inflation the
noncompact dimensions inflate whereas the compact ones are slowly changing. In
[\NIM] it was shown that enough e--foldings can be
obtained with (almost) scale invariant density perturbations of the magnitude
COBE data requires.
After asymmetric inflation ends the large compact dimensions stabilize at their
large radii.
As noted in [\NIM], in this scenario there is a cosmological moduli problem
associated with the radii of the compact dimensions, or for simplicity the
radion $r$. After early inflation ends and the large internal dimensions
stabilize the mass of the radion on general grounds (since it is a modulus)
becomes $m_r \sim M_s^2/M_P \sim 3 \times 10^{-9}~GeV$.  In addition, the
universe is reheated by inflaton decay into gravitons up to a temperature
$T_{R} \sim 10~MeV$ which is high enough for nucleosynthesis.

The generic
cosmological moduli problem manifests itself as the radion problem in this
scenario.
Following the inflationary era and reheating of the universe there is a danger
that the radion energy density will dominate the radiation energy density after
a period of time since the former decreases much more slowly than the latter.
As usual the simplest way to solve this problem is to have a second phase of
late inflation in order to dilute the radion energy density.
It turns out that in this context a late inflationary period with about 6-7
e-foldings can solve the radion problem[\NIM].

Since the radion mass is small i.e. $m_r \sim 3 \times 10^{-9}~GeV$ the late
inflation required to dilute the radion energy density has to happen at a very
small
Hubble constant $H<m_r$. In fact, there is a stronger bound on $H$ coming from
the requirement that the late inflationary period should not reintroduce
the radion problem[\CGT]. If $H$ during late inflation is too large then the
minimum of the radion
potential will be displaced by a large amount from its true value (obtained
after late inflation) and then
there will again be too much energy stored in the oscillations of the radion
field. For two large dimensions the minimum during late inflation is given by
($r_i=1$ is the minimum after late inflation)
$$r_i \sim 1+{2H^2 \over {4m_r^2}} \eqno(1)$$
On the other hand, the energy stored in the radion field is
$$V_r=2m_r^2 M_P^2 (r_i-1)^2 \eqno(2)$$
Demanding that the radion energy density does not overclose the universe today
requires that
$V_r$ sholud satisfy
$${V_r \over {T_{R}^3}} <1.5 \times 10^{-9}~GeV \eqno(3)$$
As a result, the Hubble constant during late inflation has to satisfy[\CGT]
$$H<4 \times 10^{-19}~GeV \left({T_{RH} \over {10~MeV}}\right) \left({m_r
\over {10^{-2}~eV}} \right) \eqno(4)$$
which is a constraint many orders of magnitude stronger than $H<m_r$.

In order to have late inflation the inflaton mass has to be smaller than this
value of the Hubble constant, i.e. $m_{\phi}<H$ resulting in a very small
inflaton mass. It is generically very difficult to obtain such small inflaton
masses. Moreover, since late inflation reduces the original reheating
temperature
obtained after early inflation
by an exponential factor, the decay of this light inflaton must reheat the
universe.
For such small inflaton masses it is very difficult to reheat the
universe back to $T_{R} \sim 10~MeV$.

In this letter, we show that late inflation can be realized as D--term
inflation in four dimensional compactifications of type I string theory with
D--branes (or type IIB string orientifold compactifications)[\AFIV,\IMR]. We
assume that
the compact radii are stabilized after early asymmetric inflation but before
the late D--term inflation. This is crucial for the scenario because as we will
see the very small gauge coupling required for the anomalous $U(1)$ during late
D--term inflation is a
direct result of the large compact dimensions[\EH].
 For concreteness
we consider a case of two D5 branes intersecting over three noncompact
dimensions which describe our world. One of the D5 branes is wrapped over a
small two torus
of area $M_s^{-2}$ whereas the other is wrapped over a large torus of area
$R^2 \sim M_P^2/M_s^4 \sim 2 \times 10^{17}~GeV^{-2}$. In this case, the four
dimensional gauge coupling corresponding to the interactions  on the brane
wrapping the large torus is
$$g_4^2={g_6^2 \over {R^2 M_s^2}} \eqno(5)$$
Assuming $g_6 \sim 1$ we get $g_4^2 \sim 3 \times10^{-27}$. The type I orbifold
models generically have an anomalous $U(1)$ gauge symmetry with an anomalous
D--term[\AFIV,\IMR, \NIL]. We assume that
the anomalous $U(1)$ arises from the gauge interactions on the above D5 brane
with a tiny coupling. Then there
is a D--term contribution to the scalar potential
$$V_D=g_4^2|-|\sigma|^2+M^2|^2 \eqno(6)$$
Here $\sigma$ is the trigger field which carries $-1$ charge under the
anomalous $U(1)$ and $M$ is the twisted modulus. The VEV of the twisted modulus
(which parametrizes the blowup of the orbifold singularity) $M$
fixes the vacuum energy during inflation or the Hubble constant H. In addition
we assume a tree level superpotential between the trigger field and the
inflaton $\phi$
$$W=\lambda \phi \sigma^2 \eqno(7)$$
Such tree level superpotentials are also generic to type I orbifold
models[\IMR].
Note that the inflaton is neutral under the anomalous $U(1)$. We assume that
the inflaton $\phi$ lives
on the second D5 brane wrapped over the small torus and therefore the Yukawa
coupling $\lambda \sim 1$. The superpotential gives rise to the scalar
potential
$$V_F=\lambda^2 \phi^2 \sigma^2+ \lambda^2 \sigma^4 \eqno(8)$$
The total scalar potential $V_{tot}=V_D+V_F$ gives rise to D--term inflation
(which is a type of hybrid inflation[\LIN]).

The total scalar potential has two minima; one at $\sigma=M$ and $\phi=0$ and
the other at $\sigma=0$ and $\phi$ free. Inflation occurs for the initial
conditions of large $\phi$ and small or vanishing $\sigma$. Then the vacuum
energy is dominated by the D--term and
$$V_0 \sim H^2M_P^2 \sim g^2 M^4 \eqno(9)$$
The masses of the two scalars are given by
$$m_{\sigma}^2=\lambda^2 \phi^2-2g^2M^2 \eqno(10)$$
and
$$m_{\phi}^2=\lambda^2 \sigma^2 \eqno(11)$$
For $\phi>\sqrt 2 gM/\lambda=\phi_{crit}$, $m_{\sigma}^2$ is positive and
larger than the Hubble constant, $H$. As a result, $\sigma$ rolls down to its
minimum at $\sigma=0$ very fast. At the minimum the tree level inflaton mass
vanishes.
However, there is a one--loop contribution to the inflaton mass since the
nonzero vacuum energy breaks supersymmetry
$$V_1=g^2M^4 \left(1+ {g^2 \over {16 \pi^2}}log{\lambda^2 \phi^2 \over \mu^2}
\right) \eqno(12)$$
The inflaton mass becomes
$$m_{\phi}^2={g^4M^4 \over {16 \pi^2 \phi^2}} \eqno(13)$$
Because of this small mass the inflaton slowly rolls down its potential until
$\phi \sim H$ and then inflation ends. The inflaton rolls down to its minimum
at $\phi=0$. On the other hand, $m_{\sigma}^2$ becomes negative and $\sigma$
rolls down to its new minimum at $\sigma=M$. The vacuum energy vanishes and
supersymmetry is restored.

For late D--term inflation we take the anomalous D--term scale to be the string
scale, i.e. $M \sim M_s \sim 5 \times 10^4~GeV$. Then the Hubble constant
during late inflation is $H \sim g_4 M^2/M_P \sim  10^{-22}~GeV$ which is small
enough to satisfy eq. (4). As noted earlier we need about 6-7 e--foldings to
solve the radion problem. The number of e--folds during D--term inflation is
given by
$$N \sim \left(\phi \over M_P \right)^2 {4 \pi^2 \over g^2} \eqno(14)$$
For $N \sim 6$ we need the inflaton VEV to be $\phi \sim 10^4~GeV \sim M_s$
which is its natural value. For $N$ e--foldings we find that the inflaton mass
is from eqs. (13) and (14)
$m_{\phi}^2 \sim g^2 M^4/4N M_P^2 \sim H^2/4N$.  For $N \sim 6$ we find that
$m_{\phi} \sim H/6$ and therefore the slow--roll condition is satisfied and
inflation happens. We see that the inflaton mass is very small $m_{\phi} \sim 2
\times 10^{-25}~GeV$ as required. Thus,  late D--term inflation satisfies the
requirements of late inflation such as the very small Hubble constant and
inflaton mass
to solve the radion problem.

Note that during late D--term inflation the magnitude of density perturbations
is very small
$${\delta \rho \over \rho} \sim {\lambda g^2 M^5 \over {M_P^3 m_{\phi}^2}}
\eqno(15)$$
giving $\delta \rho /\rho \sim 10^{-13}$ which is completely negligible. This
is fine because the density perturbations of the early inflationary era were of
the corect magnitude and (almost) scale invariant. Therefore, the magnitude of
the density
perturbations satisfies COBE constraints for the period of late inflation.

Finally as mentioned in [\CGT] a very light inflaton as above cannot in general
reheat the universe up to the MeV scale. However, note that after D--term
inflation ends the inflaton mass becomes much larger than what it was during
inflation. After D--term inflation
$$m_{\phi}^2 \sim \lambda^2 M^2 \sim 3 \times 10^9~GeV^2 \eqno(16)$$
The main decay channel for the inflaton is to two fermionic $\sigma$'s due to
the large coupling coming from the tree level superpotential in eq. ().
The decay rate of the inflaton into two fermionic $\sigma$'s is given by (note
that the mass of $\sigma$ after inflation vanishes since $\phi=0$)
$$\Gamma \sim {\lambda^2 m_{\phi} \over {8 \pi}} \sim 2 \times 10^3~GeV
\eqno(17)$$
This is much larger than the Hubble constant, $H \sim 10^{-22}~GeV$ and
therefore all the  vacuum energy will be converted into heat very efficiently.
As a result,
$$T_R^4 \sim g^2 M^4 \eqno(18)$$
and $T_{R} \sim \sqrt{g} M \sim 10^{-2}~GeV$.
This is about the normalcy temperature required in order to reheat the brane
and not
predominantly the the bulk by graviton production[\EXP]. It is also above the
nucleosynthesis scale as required for maintaining the success of late stages of
cosmology.

\bigskip
\centerline{\bf 3. Conclusions}
\medskip

In this letter we have shown that late D--term inflation can solve the
cosmological moduli problem which arises in inflationary models at the TeV
scale.
In order for this to be possible the Hubble constant and the inflaton mass must
be extremely small. This requires the large dimensions to be stabilized before
late D--term inflation. This
is a reasonable assumption since the Hubble constant during late inflation is
extremely small and therefore it happens at very late times. The size of the
large compact dimensions is responsible for the very small coupling of the
anomalous $U(1)$ which in turn results in the extremely small Hubble constant
and inflaton mass required. Even though the inflaton mass (during late
inflation) is extremely small its decay into fermionic $\sigma$'s can reheat
the universe to the $MeV$ scale because the
inflaton mass after inflation is many orders of magnitude larger. This is one
of the important properties of D--term inflation (or actually hybrid
inflation).

The proposed solution of the cosmological moduli problem by late D--term
inflation is more general than the above example. In particular, the modulus
does not have to be the radion and
the early phase of inflation does not have to be asymmetric inflation.
If one
assumes that the compact radii are fixed before early inflation then this
can be realized by D--term inflation[\EH,\HBD]. Since D--term inflation is
realized in the context of type I string (or type IIB orientifold)
compactifications
there will be string moduli such as the dilaton, untwisted and twisted moduli.
(For the phenomenology of these see[\HAL].)
These generically are expected to have very small masses of $O(10^{-9}~GeV)$
which they obtain after supersymmetry breaking around $M_s$ on the brane. As
usual, there will be a
cosmological moduli problem associated with them after early D--term inflation
with $H \sim 10^{-10}~GeV$ ends and the universe is reheated. The amount of
e--foldings required to solve this problem is exactly what was needed to solve
the radion problem of asymmetric inflation, about 6-7 e--foldings. Once again
we need late inflation with a very small Hubble constant at least smaller than
the moduli masses. However, in the previous section we saw that for the radion
a much stronger bound on $H$ was obtained from the requirement that late
inflation should not introduce the cosmological moduli problem. The bound was
obtained
from the detailed knowledge of the radion minimum during inflation. This is not
in general possible for string moduli but it is safe to assume the same bound
on $H$ in this case too.

In this scenario there are two anomalous gauged $U(1)$ symmetries with vastly
different scales, e.g. $M_1 \sim 10^{10}~GeV$ and $M_2 \sim 10^4~GeV$
which refer to the scales of early and late D--term inflation. Equivalently,
these two scales refer to the VEVs of two twisted moduli which parametrize
blowups of different orbifold singularities. It is interesting to note that
type I string orbifold models generically have more than one anomalous gauged
$U(1)$'s[\AFIV,\NIL]. However we do not have an expalnation for the two vastly
different scales.

Above we mainly concentrated on the case with two large compact dimensions.
For more than two large dimensions the only difference is in the string scale.
In these cases the bound on the string (or higher dimensional Planck) scale is
weaker, i.e. $M_s>1~TeV$. This increases the modulus (or radion) mass. In
addition, the bound on the Hubble constant during late inflation becomes much
weaker, e.g. from $O(10^{-19}~GeV)$ up to $O(10^{-14}~GeV)$ depending on the
number of large dimensions.
In any case, it is clear that all our results apply equally well to the cases
with more than two large dimensions.

\bigskip
\centerline{\bf Acknowledgements}

We would like to thank Nemanja Kaloper for raising the possibility of late
inflation and for very useful discussions.

\vfill

\refout

\end
\bye